# A deep learning approach for predicting the quality of online health expert question-answering services

Ze Hu[a], Zhan Zhang[a], Qing Chen[b], Haiqin Yang[c], Decheng Zuo[a, *]

[a]School of Computer Science and Technology, Harbin Institute of Technology, Harbin 150001, China

[b]Research Center on Satellite Technology, Harbin Institute of Technology, Harbin 150001, China

[c]Department of Computing, Hang Seng Management College, Hong Kong

*Corresponding author e-mail: zuodc_hit@163.com

# ABSTRACT

Currently, a growing number of health consumers are asking health-related questions online, at any time and from anywhere, which effectively lowers the cost of health care. The most common approach is using online health expert question-answering (HQA) services, as health consumers are more willing to trust answers from professional physicians. However, these answers can be of varying quality depending on circumstance. In addition, as the available HQA services grow, how to predict the answer quality of HQA services via machine learning becomes increasingly important and challenging. In an HQA service, answers are normally short texts, which are severely affected by the data sparsity problem. Furthermore, HQA services lack community features such as best answer and user votes. Therefore, the wisdom of the crowd is not available to rate answer quality. To address these problems, in this paper, the prediction of HQA answer quality is defined as a classification task. First, based on the characteristics of HQA services and feedback from medical experts, a standard for HQA service answer quality evaluation is defined. Next, based on the characteristics of HQA services, several novel non-textual features are proposed, including surface linguistic features and social features. Finally, a deep belief network (DBN)-based HQA answer quality prediction framework is proposed to predict the quality of answers by learning the high-level hidden semantic representation from the physicians' answers. Our results prove that the proposed framework overcomes the problem of overly sparse textual features in short text answers and effectively identifies high-quality answers.

# 1. Introduction

Globally, aging problems are becoming increasingly prominent. When aging is accompanied by higher rates of chronic diseases, such as hypertension, diabetes, cancer and cardiac disease, the result is severe shortages of medical resources such as physicians, nurses, hospital rooms and health educators [1]. Moreover, the distribution of these medical resources is also extremely unbalanced, as most well-qualified physicians and hospitals are typically located in large cities and developed areas. To secure the most qualified and professional treatment possible, health consumers must spend large amounts of time and money, leading to an enormous waste of social resources. Additionally, rising health care costs have become an increasingly pressing social issue [2]. With the rapid growth of the Internet, many health consumers now tend to search for health information online [3-9]. The most convenient and economical method for this purpose is to use a search engine or a question-answering service for health consultation [10-13]. In 2012, 59% of adults in the US searched for health information online. This ratio is even higher in large cities [14]. This green and environmentally friendly health consultation method effectively alleviates medical resource shortages and the lack of balance in the distribution of medical resources. A search engine generally takes health query keywords from health consumers as input to retrieve a large volume of typically redundant and non-comprehensive archives, and health consumers must then, based on their own experience, spend time identifying the answers that best meet their requirements. An online question-answering service provides a communication platform specifically for health consumers[15]. Using such a platform, health consumers communicate and share health knowledge in the form of question answering (QA). In general, there are two types of QA services: community-based QA (CQA) and expert-based QA (EQA) services. In CQA, any user can ask other community members questions about any topic, answer questions, and rate or comment on answers [16-20]. In EQA, only qualified experts can answer questions, while common users can only ask questions. Compared with CQA, services that offer EQA for health questions (abbreviated as HQA in this paper) provide health consumers with more trustworthy, more professional and higher-quality answers from reliable sources. Mobile HQA applications enable health consumers to initiate health consultations with physicians at any time and from anywhere[21], thereby to some extent accomplishing patient self-health management.

HQA provides health consumers with high-quality health-related answers; however, current practices still have certain deficiencies. First, not all answers are of high quality. For instance, the answers from some physicians are merely advertisements for particular hospitals or other irrelevant information. These low-quality answers cannot actually solve health consumers' problems, and the consumers must continue to consult other physicians until they receive satisfactory answers. This process severely affects the health consumers' user experience and the HQA service's reputation, wastes medical resources and lowers efficiency. Second, the questions from health consumers tend to be conversational in tone, and there may be a large number of repeated or similar questions. A health question-answering system based on an existing HQA knowledge base can provide health consumers with timely and effective answers [22-25]. However, low-quality answers result in underutilization of the existing HQA knowledge base, requiring health consumers to consult with physicians once again, thereby leading to repeated effort on the part of the physicians and lowering efficiency. Therefore, it is of great importance to automate the evaluation and prediction of the quality of existing answers to a problem.

In current CQA service answer-quality prediction, a common practice is to extract a group of features (including textual and non-textual features) from a dataset, which is then used as input for a classification system to determine answer quality. It is worth noting that, as CQA services include community attributes such as best answer and user votes, the "wisdom of the crowd" can be leveraged to define answer quality. Therefore, when predicting CQA service answer quality, it is normally assumed that the answer chosen by the questioner or with the highest number of votes is a high-quality answer. However, HQA services have no community attributes such as best answer and user votes. Therefore, the "wisdom of the crowd" is unavailable to assess answer quality. Furthermore, every question in a CQA service receives one or more complete candidate answers, whereas each QA pair in an HQA service is just a small part of a complete physician-patient interaction. Therefore, the answers in an HQA service are normally short texts with incomplete information, resulting in insufficient context information and a high likelihood of being severely affected by the data sparsity problem. In addition, as HQA services and CQA services are different in nature, non-textual features commonly used in CQA service answer quality prediction are inapplicable to HQA services. Therefore, how to define and predict answer quality in an HQA service is a difficult and challenging task.

To address these problems, we focus on the task of predicting HQA answer quality. Inspired by related research on CQA answer quality evaluation [26-30], HQA answer quality prediction is defined as a classification task. In other words, features, including both textual features and non-textual features, are extracted from an HQA knowledge base and then used as input to a classification system to predict answer quality.

In the case of textual features, the answers provided in the HQA context are typically in the form of short texts. The text length limitation leads to a sparsity of textual features based on word frequency. To overcome this problem of overly sparse textual features, a deep belief network (DBN)-based [31-38] deep learning method is proposed for the extraction of textual features. The proposed method is compared with three common textual feature extraction methods for CQA answer quality prediction: the most common high-frequency-word-based textual feature extraction method, which considers only keyword frequency; a CHI-TFIDF-based textual feature extraction method, which considers keyword frequency and the class differentiation of keywords [39]; and a textual feature extraction method based on Latent Dirichlet Allocation (LDA) [40], which considers the topic distribution of the text. The test results show that when predicting the quality of HQA answers with a relatively short text length, the DBN-based textual feature extraction method is significantly superior to these conventional textual feature extraction methods.

With regard to non-textual features, the similarities and differences between CQA and HQA are compared and analyzed. It is shown that several non-textual features that are closely related to the CQA website structure and are commonly used in related studies concerning CQA answer quality evaluation, such as the number of times that an answer is printed, the number of recommendations by editors, and the similarities between answers [27], are not applicable to HQA answer quality evaluation. By contrast, several non-textual features that are unrelated to the structure of the question-answering service website, such as the answer length, word density, similarity between a question and the corresponding answer, length ratio between a question and the corresponding answer, number of sentences, and number of high-frequency domain words [17, 20, 26-30, 41-43], are still applicable to HQA answer quality evaluation. Therefore, several general features of related studies concerning CQA answer quality prediction are reused, and several novel features based on the HQA website structure are proposed. The proposed non-textual features are divided into two categories: surface linguistic features and social features. The intuition for the proposed surface linguistic features is as follows: surface linguistic features contain features that are based on statistics over answer content, which approximately reflect the attributes of the answer's writing style, including fluency, grammar and expression [44]. For instance, the number of unique words in the answer may reflect fluency. These features also approximately reflect the answer's professionalism and authority. For instance, the number of high frequency domain words and the keyword density may reflect the authority of an answer. The surface linguistic features also include the statistical features of the QA pair, which approximately describes the relationship between the question and answer. For instance, the number of overlapping words in the QA pair and the similarity between the question and answer may reflect the relevance between the answer and question. Therefore, we believe that surface linguistic features are relevant to HQA service answer quality prediction. The intuition for the proposed social features is as follows: Zhang et al. discovered that users' personal information is very effective in answer rating tasks, helping to quickly identify community experts [45]. Thus, an expert or a user with professional knowledge on a particular topic is more likely to provide a high-quality answer [42]. Proposed social features include the physician's profile. Since the profile reflects user behavior and historical statistics, similar to the answer rating task, we believe that the proposed social features are also applicable to HQA service answer quality prediction. The test results show that the proposed features and reused general features can effectively determine the quality of HQA answers.

Furthermore, a DBN-based HQA answer quality prediction framework is proposed. This framework includes two operational phases: first, sparse textual features based on word frequency are used as input to a deep learning network to learn a non-sparse high-level hidden semantic representation; then, this high-level hidden semantic representation learned by the network and the relevant non-textual features are integrated into a unified representation, which is used as input to a classifier to train an answer quality prediction model. A large number of test results prove that the proposed framework and the proposed novel non-textual features effectively improve the performance of HQA answer quality prediction.

The main contributions of this paper can be summarized as follows:

1) It is the first time that the evaluation and prediction of the quality of HQA service answers have been automated.
2) A set of quality evaluation standards is proposed to define answer quality in an HQA service.
3) A DBN-based deep learning approach is proposed to create a model for the high-level semantic representation of physician answers in an HQA service. Based on word occurrence features alone, the high-level semantic representation learnt by this model demonstrates excellent performance in predicting answer quality. Based on this data-driven strategy, semantic knowledge is learnt from a large number of physician QA pairs, which effectively overcomes the feature sparsity problem in short text answers.
4) Some novel non-textual features are proposed, and the quality of a physician's answer is

described from two angles: surface linguistic features and social features. Additionally, surface linguistic and social features, their extraction methods, and the intuition and motivation to propose these features are elaborated. The distributions of some of the most important surface linguistic features and social features in the dataset are analyzed in detail, and why the proposed non-textual features are effective for the HQA service answer quality prediction task is investigated. Moreover, the significance of non-textual features is analyzed to identify which features effectively represent a high-quality answer.

5) The performance of various combinations of textual features and non-textual features is assessed. The results show that the combination of one textual feature set and two non-textual feature sets significantly improves the performance of answer quality prediction. In particular, the proposed deep learning approach achieves optimal prediction performance by almost every evaluation metric. This finding sufficiently proves the effectiveness of the proposed approach and features.

6) Thorough experimental analysis is performed for the proposed deep learning-based approach and novel non-textual features. All test results were obtained after averaging the results of 5 trials and performing 5-fold cross-validation. The standard deviations for all results are almost always under 3%, which effectively proves the stability of the test results. Moreover, the T-test is performed for comparison with existing methods, and all $p$ values are under 0.01, which shows that the proposed deep learning-based approach is stable, reliable and significantly superior to existing methods for the HQA answer quality prediction task.

The rest of this paper is organized as follows: section 2 gives a brief overview of relevant research; section 3 introduces a standard definition for answer quality, a definition of the problem, the modeling process of the feature learning module, the extraction of non-textual and textual features; section 4 reports the test results and detailed analysis; and finally, section 5 presents the conclusions and proposed future work.

## 2. Related work

Our research primarily focuses on how to extract a series of textual features and non-textual features from an HQA service dataset and how to use these features in a classification system to predict answer quality. Since an answer in an HQA service is normally a short text, our research involves short text classification. As our research is essentially a text quality prediction task, our research naturally also involves text quality prediction. Moreover, the intuition regarding the social features proposed in section 1 shows that an expert or user with professional knowledge in a certain area is more likely to provide a high-quality answer. Therefore, our research is also related to how to identify community experts. Basically, existing works related to our research include three topics: short text classification, text quality prediction and community expert identification.

### *2.1. Short text classification*

In studies on short text classification, a common practice is to introduce extra features via various methods to extend the short text and overcome the data sparsity problem in short text representation. Phan et al. obtained a series of latent topic features from the Wikipedia corpus via LDA and extended short texts via the topic features obtained from the proposed general framework [46]. Yan et al. improved LDA and proposed the bi-term topic model (BTM), which effectively overcomes the data sparsity problem in short text modeling [47]. Sahami and Heilman used short texts as the input for a search engine and extended the short texts with the results returned [48]. Zhou et al. introduced extra semantic information from Wikipedia to extend short texts to improve question retrieval performance in a CQA service [49]. Chen et al. discovered that multiple granularity topic learning could create a more accurate short text model and improve short text classification performance [50]. Kim et al. introduced extra semantic knowledge via pre-trained word embedding from Google, and they classified short text sentences via a dual-channel convolution network [51].

*2.2. Text quality prediction*

Text quality prediction primarily includes webpage quality prediction and answer quality prediction. For the webpage quality prediction task, there are two mainstream methods: the link analysis [52-54] and feature methods. The latter is closer to our research. The feature method is widely used in webpage quality prediction tasks. For instance, Zhu and Gauch manually marked webpage documents as one of three categories: good, normal and inferior; these features were used to train a webpage quality prediction model that predicted the quality of other webpage documents with unknown quality levels [55]. For the answer quality prediction task, the majority of research primarily focuses on CQA service answer quality. In a CQA service, anyone can ask or answer questions. Due to the varying levels of user expertise, the answer quality varies considerably from extremely high to extremely low. There may even be advertisements or abuses. Therefore, evaluating and predicting CQA service answer quality is a challenging task. A common solution to this challenge is to extract a series of textual features and non-textual features from a dataset and use them as input for a classification system to predict answer quality. Jeon et al. predicted answer quality for the first CQA service in the world, Naver (http: //www.naver.com/). They extracted non-textual features based on QA pair context information and then combined these features with a maximum entropy model to create an answer quality prediction framework [27]. Later, Agichtein et al. and Bian et al. proposed a richer feature set including community features, textual features and structure features. All these features were used as input for a classification framework to identify high-quality answers in a CQA service [26, 56]. Shah and Pomerantz extracted some non-textual features (mostly meta-information) from Yahoo! Answers (http: //answers.yahoo.com/). This information was combined with 13 different quality evaluation standards to create a classification method for answer quality evaluation and prediction [30]. Cai et al. proposed a set of novel time sequence features and combined them with a learning to rank model to improve the answer quality prediction for a CQA service [57]. Finally, Harper et al. investigated whether a pay system was more likely to provide high-quality answers [58]. The above research shows that how to fully utilize textual features is often ignored in answer quality prediction for CQA services. This omission may be because the answer text is too short to extract valid features, preventing the effective improvement of answer quality prediction.

### 2.3. Community expert identification

Some researchers believe that identifying expert users in CQA services is very helpful for tasks such as CQA service answer quality evaluation and question retrieval and sorting, as they assume that a CQA service user with a higher level of authority normally provides higher-quality answers. Zhang et al. proposed the ExpertiseRank algorithm to identify expert users in CQA services [45, 53]. Jurczyk et al. applied the HITS [52] algorithm to a user-answer graph in an online forum to calculate user authority [59]. Dom et al. and Campbell et al. discovered that in the community expert identification task, a link-based algorithm had better performance than a content-based algorithm [60, 61]. In the aforementioned work, the researchers discovered a strong correlation between expert user networks and answer quality. Furthermore, Bian et al. combined textual content and other community features and, based on a mutual reinforcing relationship, proposed a ranking algorithm that could be applied to a variety of networks consisting of various node types such as user, question and answer [56].

## 3. Materials and methods

### 3.1. Answer quality characterization

To provide users with high-quality answers, existing CQA services normally employ four methods to control user answer quality: the expert selection method, Wikipedia method, user vote-based method and questioner satisfaction method [57]. The expert selection method requires experts to pass a strict qualification certification before they are allowed to provide answers to users. Each question is only answered by an expert. An example of a community with such quality control is AllExperts (http: //www.allexperts.com/). In the Wikipedia method, each question only has one answer; however, this answer can be updated and improved by multiple users. In this way, the questioner is presented with a high-quality answer that includes the wisdom of the crowd. An example of a community with such quality control is Answers (http: //wiki.answers.com/). In the user vote-based method, a question normally has multiple answers; each answer is voted for by different users, and the answer with the highest number of votes is normally regarded as the answer with the highest quality. This quality control method is the most widely adopted method in CQA service, and a typical example is Yahoo! Answers (http: //answers.yahoo.com/). In the questioner satisfaction method, a question normally has multiple answers, and the questioner determines the quality of an answer. If the questioner is satisfied with an answer, then this answer is marked as the best answer. Although this method is rather subjective and problematic, it is commonly used in research on predicting CQA service answer quality to determine answer quality. In other words, it is assumed that the best answer marked by the questioner is a high-quality answer, while the other answers are low-quality answers. Yahoo! Answers (http: //answers.yahoo.com/) also uses this method of quality control. It is worth noting that many CQA services use more than one quality control method. For instance, Yahoo! Answers (http: //answers.yahoo.com/) uses both the user vote-based method and the questioner satisfaction method.

Quality control in an HQA service is similar to the expert selection method, requiring a physician to pass a qualification certification before answering questions from patients. However, the current practice has severe deficiencies, and the expert selection method is almost ineffective in HQA services. The expert selection method only guarantees that physicians are capable of providing high-quality answers when they apply for qualification. However, there is no guarantee that a given physician actually provides high-quality answers after certification. Whether a physician provides high-quality answers after certification depends entirely on the physician's attitude and expertise. Analysis of the dataset actually confirms this point. For instance, due to limitations of professional knowledge, a physician is unlikely to provide a high-quality answer when a patient asks a question beyond his expertise. Alternatively, a well-known physician is often overloaded with a huge number of questions from patients and has inadequate time to provide high-quality answers to every patient. Moreover, some less well-known physicians post advertisements everywhere to improve their rankings without effectively solving the patient's problem. Therefore, how to define HQA service answer quality and automate HQA service answer quality evaluation and prediction remains an enormous challenge. To address this challenge, based on the characteristics of the HQA service dataset sample and suggestions from medical experts, a set of standards by which to evaluate HQA service answer quality is proposed. Answers by physicians with the following features are regarded as low-quality answers:

1) Reply provides little help to patient;
2) Reply provides little relevant guidance or suggestions but often includes personal contact information;
3) Clinic office is moderate in expertise level but always promotes a certain technology or treatment method to the patient or labels itself as top level;
4) Immature treatment technologies, such as diabetes treatment via stem cell transplants, are promoted;
5) Physician always suggests the patient upload a photo;
6) Physician always suggests the patient visit the hospital directly;
7) Physician always states that the question is overly simple;
8) Physician always replies using a template;

9) Physician always promotes self and the hospital the physician works in;

10) Physician has an extremely large number of offline outpatients;

11) Reply is of some help to the patient but is overly simplified, hindering subsequent communication.

By contrast, answers with the following features are regarded as high-quality answers:

1) Physician with low grade provides conclusive suggestions highly relevant to patient's question plus a simple explanation;

2) Physician with high grade provides conclusive suggestions highly relevant to patient's question plus a detailed explanation.

## 3.2. Problem Definition

Based on the quality standards in section 3.1, the HQA service answer quality prediction problem is defined as follows: given a question and an answer for this question, the goal is to calculate answer quality via the proposed deep learning-based method and novel non-textual features, that is, to calculate the probability of an answer being in the high-quality category. If the probability is above 0.5, then this answer is regarded as a high-quality answer; otherwise, it is a low-quality answer. Moreover, no special optimization is applied in the deep network training, as the purpose is to prove that even in the worst-case scenario, our method and features can still achieve stable and superior performance. With the assistance of the proposed framework for automated HQA service answer quality prediction, the HQA service effectively identifies the quality of a physician's answer, and the result is close to human judgment. Therefore, it is possible to identify and block low-quality answers in a timely manner to improve overall user satisfaction.

## 3.3. Dataset and evaluation metrics

The test dataset was based on the largest and most popular HQA service in China: Haodf Online (http: //www.haodf.com/). On this service, questions are asked by health consumers in China and answered by physicians in China. Over 45216 resolved problems were collected from Haodf Online, covering 29 level one categories, with a time span from June 2008 to March 2016. To filter out noise from the original dataset, a criterion for question selection was defined to require that each answer should contain at least 15 characters. In this way, 15206 questions, 15206 answers, and 6540 physicians were selected. Based on the physicians' names, the profile web pages of all the physicians were crawled to collect social features. Because Haodf Online does not offer attributes that allow selection of the best answer or user voting, the "wisdom of the crowd" was unavailable for automatically marking the quality of the physicians' answers. Therefore, the physicians' answers were manually marked. In the marking process, a quality label for high-quality and low-quality ratings was employed. Three medical professionals spent 90 days marking each QA pair independently based on evaluation metrics in section 3.1 and related work [11]. For each QA pair, when at least two of the medical professionals marked it as high quality, the final tag for that QA was high quality, and the same was true for marking as low quality. After marking, the corpus contained approximately 89% high-quality answers and 11% low-quality answers. It is assumed that most physicians will tend to provide high-quality answers, and thus, an HQA knowledge base will typically contain more high-quality answers than low-quality answers. To provide health consumers with the best user experience, it is vital for an HQA service to maintain overall performance while promptly identifying and discarding low-quality answers. Because the collected data were inherently imbalanced, to prevent any prediction deviation originating from the original data, 1600 QA pairs were extracted randomly from high-quality QA pairs as the positive category, and 1600 QA pairs were extracted randomly from low-quality QA pairs as the negative category. In each category, 80% of the QA pairs were chosen randomly as the training set, and 20% of the QA pairs were used as the test set. All test results were obtained after averaging the results of 5 trials and performing 5-fold cross-validation.

In this paper, because the task of HQA answer quality prediction is defined as a two-class classification issue, the precision (P), the recall (R), the Fl-measure (Fl), and the Area Under the ROC Curve (AUC) are used as the main evaluation metrics.

### 3.4. Framework for HQA answer quality prediction

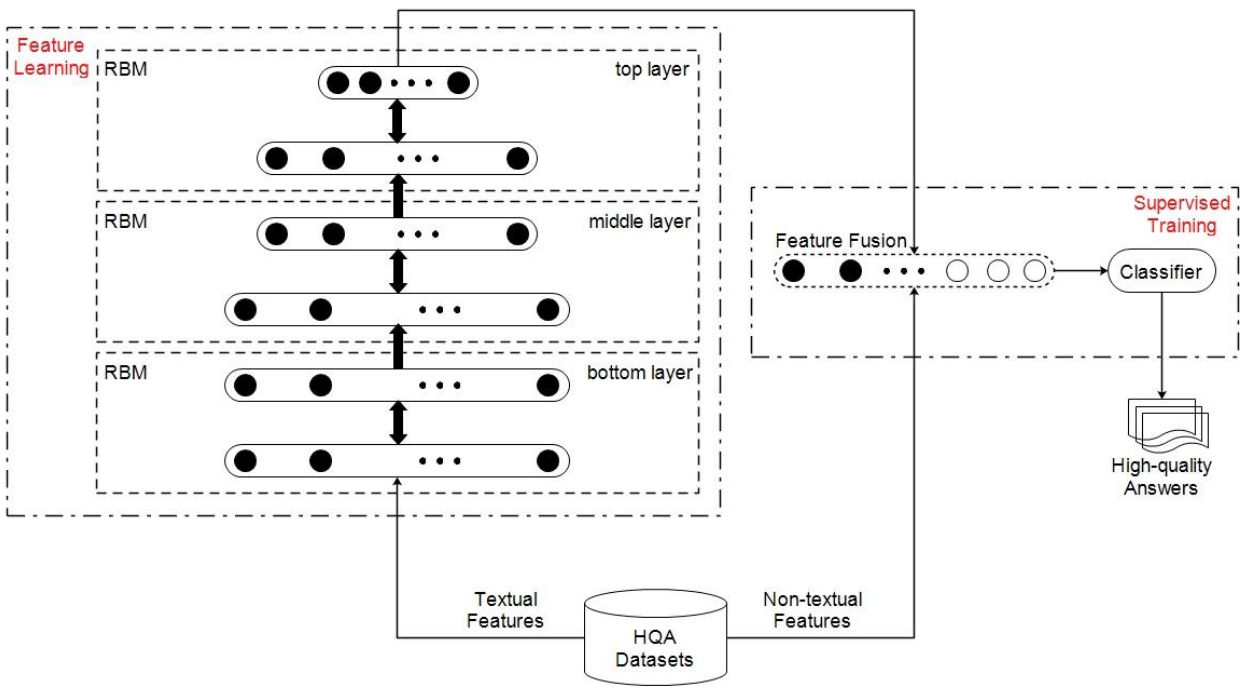

Fig. 1. DBN-based framework for HQA answer quality prediction.

HQA answer quality prediction is defined as a typical classification issue. Fig. 1 shows an overview of the proposed framework for DBN-based HQA answer quality prediction. First, textual features are extracted from the HQA dataset based on the most frequent words and weighted using a binary scheme. These features are used to train a DBN model and to learn a high-level hidden semantic representation of the physicians' answers. Next, this high-level hidden semantic representation learned by the deep learning architecture and the alternative perspective provided by the relevant non-textual features, including the surface linguistic features and social features, are integrated into a final unified representation. Finally, this unified representation is used as the input to train a classifier to predict answer quality.

As mentioned previously, in the HQA context, the physicians' answers are typically provided in the form of short texts, and the limitation on the text length leads to overly sparse word-based textual features. Moreover, the word frequencies in short texts usually take values of only 0 or 1; in other words, aside from whether a word is present in the text, the word frequency provides extremely limited information. Therefore, in studies on CQA, many researchers have attempted to introduce non-textual features such as structural features [27, 62] or user behavioral features [42] to improve the performance of answer quality prediction models, and the effects of textual features are ignored. However, this approach has severe limitations, as a user who typically provides high-quality answers may sometimes provide low-quality answers for various reasons, whereas users who typically provide low-quality answers may occasionally provide high-quality answers under certain circumstances. Relying on non-textual features alone for answer quality prediction will result in a certain degree of prediction deviation; therefore, it is preferable to fully utilize the available textual features to improve model prediction performance. For this reason, a DBN-based model is proposed in this study to extract the semantic information hidden in sparse word-frequency-based textual features. The test results demonstrate the powerful capability of this DBN-based method. Even when only word-frequency-based textual features are used, the proposed method is already superior to conventional methods of answer quality prediction.

### 3.5. Restricted Boltzmann machines

A "restricted Boltzmann machine" (RBM) [63-65], as a two-layer random neural network, effectively overcomes the problems of a Boltzmann machine (BM) [66], such as its inability to accurately calculate the represented distribution and its long training time [67]. An RBM-based dimension reduction method facilitates the storage, classification, exchange and visualization of high-dimensional data [31]. An RBM-based deep graphical model enables the execution of semantic hashing to allow similar documents to be retrieved faster and more accurately [68].

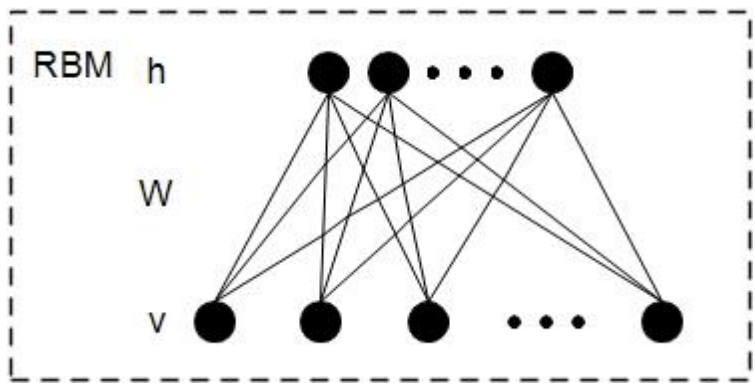

Fig. 2. Restricted Boltzmann machine.

In Fig. 1, the feature-learning component of the proposed framework is a DBN consisting of three RBMs. As shown in Fig. 2, an RBM is a two-layer undirected graphical model with symmetric connections and without self-feedback. In this model, **v** denotes the stochastic visible units, which represent the observed data; **h** denotes the stochastic hidden units, which are regarded as feature detectors; and the symmetric matrix **W** provides the connection weights between the two layers of the RBM. For a given input vector **v**, after training, the RBM generates a hidden feature vector **h**, which is used to reconstruct **v** with minimum error. Classical RBMs have been effectively applied for the modeling of binary data distributions, whereas Gaussian RBMs can effectively model real-valued data [69]. Testing shows that an RBM provides an effective means of feature extraction and that a DBN consisting of multiple RBMs can extract more abstract features. This special capability of RBMs suggests that a suitably constructed RBM-based DBN can effectively model the semantic information hidden in physicians' answers.

## 3.6. Feature learning

As shown in Fig. 1, the feature-learning component includes a DBN model. The model consists of three layers, each of which is an RBM. The bottom layer is used as an example to illustrate the feature-learning process. The bottom layer generates hidden features from the visible answer vector of the training set, which are then used to reconstruct the answer vector and learn the semantic information hidden in the physicians' answers. This reconstruction process converges. At the bottom layer, the word-frequency statistics-based binary feature vectors of the answers are used as input for the visible layer in the RBM. These visible feature vectors are employed to calculate the hidden features for the hidden layer. The RBM reconstructs the answers from these hidden features. Inspired by the work of Hinton and Salakhutdinov [31, 68], the process is modeled as follows:

$$p(h_j = 1 | \mathbf{a}) = \sigma(b_j + \sum_i a_i W_{ij}) \quad (1)$$

$$p(a_i = 1 | \mathbf{h}) = \sigma(b_i + \sum_j W_{ij} h_j) \quad (2)$$

where $\sigma(x) = (1 + \exp(-x))^{-1}$ is a sigmoid activation function, $\mathbf{a}$ is the visible feature vector of the answers, $\mathbf{h}$ is the hidden feature vector for answer reconstruction, $a_i$ is the state of word $i$ in the answer vector, $h_j$ is the $j$-th hidden feature, $W_{ij}$ is the connection weight between word $i$ in the answer vector and the hidden feature $j$, $b_i$ is the model bias for word $i$, and $b_j$ is the bias for the hidden feature $j$.

The original answer vectors are used as the input to initialize the states of all visible units in the bottom layer. The binary states of all hidden units are calculated via Eq. (1) to generate the corresponding hidden features. After the hidden features are stochastically activated, the Bernoulli rate for each word in the answer vector is reconstructed via Eq. (2) for subsequent answer vector reconstruction. Next, Eq. (1) is employed again to activate the hidden features. The parameters are updated by executing stochastic gradient ascent and the one-step Contrastive Divergence algorithm [63]:

$$\Delta W_{ij} = \epsilon(<a_i h_j>_{aData} - <a_i h_j>_{aRecon}) \quad (3)$$

where $\epsilon$ is the learning rate; $<a_i h_j>_{aData}$ is the expected answer data distribution, which defines the frequency with which word $i$ in the answer vector will coexist with feature $j$ when the hidden features are driven by the answer data in the training set; and $<a_i h_j>_{aRecon}$ is the expected model-defined distribution, i.e., the corresponding expectation when the hidden features are driven by the reconstructed answer vectors. The parameter update rule for the biases is merely a simplified version of Eq. (3).

The training method for the middle and top layers is identical to the training method for the bottom layer. After one layer is trained, the activation probabilities of hidden units are used as training sets for the next higher-level layer. It is worth noting that the function of each layer is to decrease the dimension of the answer vectors. Different layers have very similar structures, except for slightly different numbers of visible units and hidden units. The features of the RBMs ensure the generation of hidden features that sufficiently capture the semantic information contained in the answer vector. Following the steps described above, a deep learning network is created to learn a high-level hidden semantic representation from the textual features of the original answers.

### *3.7. Supervised training and classification*

Once high-level hidden semantic representation is learned via the feature-learning component, as depicted in Fig. 1, that representation is merged with non-textual features extracted from different perspectives to form a unified representation. Next, this unified representation is used as the input for classifier training. Several common classifiers, including Logistic Regression (LR) [70], Support Vector Machines (SVM) [71], Factorization Machines (FM) [72], and Naïve Bayes (NB) [73], were employed to estimate the effectiveness of the presented method and features. The LR classifier was employed for the final prediction because it exhibited the best classification performance.

*3.8. Textual features*

In areas such as natural language processing, text mining, information retrieval and machine learning, the most common representation method used for text documents is the bag-of-words model. In the bag-of-words model, text documents are treated as an assembly of high-frequency words, and all words in the text document are assumed to be independent. Although the bag-of-words model is simple and intuitive, it is unable to effectively capture word-level synonymy and polysemy [74]. In particular, for short text documents such as the answers provided in HQA, the bag-of-words model encounters the problem of severe feature sparsity. Previous studies have also proven that the bag-of-words model cannot be used for high-quality content recognition [26]. In this study, two types of feature weight calculation methods were employed to construct two different bag-of-words models as baselines. For the first baseline, after the usual pre-processing steps (i.e., word segmentation, part-of-speech tagging, and removal of stopwords), 1904 high-frequency words were extracted from the training set to construct a bag-of-words model. Next, each answer was represented in the form of a binary weighted vector of 1904 dimensions. For the second baseline, the CHI-TFIDF textual feature extraction method [39] was employed after the usual pre-processing to extract 1904 features from the training set to construct a bag-of-words model considering not only keyword frequency but also the contribution of each keyword category. Afterward, each answer was represented in the form of a TFIDF [75] weighted vector of 1904 dimensions. In addition, a topic model was employed to extract textual features because topic models can represent text documents at a coarser granularity and to thus generate a more powerful representation of text documents that can effectively neutralize feature sparsity to some extent. In this paper, the topic model is treated as the third baseline method, and a typical LDA topic model was employed in the test. LDA can be used to extract the topic distribution of all words. Then, based on the extracted topic distribution of all words, the topic distribution for a new text can be deduced. Topic model features are suitable for predicting the answer quality in HQA. In this case, each answer was represented in the form of a vector of 25 topics weighted by topic distribution. These 25 topics were determined in two steps: first, based on the scale of the dataset and the experience of the analyst, an approximate topic range was estimated; then, within the estimated topic range, multiple classification tests were performed to identify the topics with the optimal classification performance. Finally, the proposed DBN-based deep learning approach was used to extract abstract textual features. For details, please refer to the section 3.6. In this paper, the deep learning approach is treated as the fourth baseline method. To construct the DBN for the feature-learning component of the framework shown in Fig. 1, a 1904-1904-1500-1000 architecture was employed. Thus, the bottom layer contained 1904×1904 elements, the middle layer contained 1904×1500 elements, and the top layer contained 1500×1000 elements. Based on this DBN, a binary vector of 1904 dimensions could eventually be mapped to a real-value vector of 1000 dimensions. In the DBN training phase, the bottom layer underwent 50 rounds of greedy training on the entire training set, and the other two layers also each underwent 50 rounds of greedy training. In each training round, the parameters for each layer were updated via the one-step Contrastive Divergence algorithm [63]. During the DBN training process, to prevent overfitting, the weight cost was set to a small value of 0.0002, and the learning rate for each layer was set to 0.6. To accelerate DBN training, momentum technology was employed. In the first five epochs, the momentum was set to 0.5; in the remaining epochs, the momentum was set to 0.9. For further details about the deep training architecture, please refer to the literature [76].

### 3.9. Non-textual features

In CQA, the majority of the necessary work on how to leverage non-textual features for user answer quality prediction has been completed. Jeon et al. proposed an answer quality prediction framework based on non-textual features and a maximum entropy model and used this framework to predict user answer quality for the first CQA in the world, Naver (http: //www.naver.com/) [27]. Later, Bian et al. and Agichtein et al. proved that structural features and community features could effectively identify high-quality answers in the CQA context [26, 56]. In addition, Shah and Pomerantz proposed 13 quality evaluation criteria to identify useful features for evaluating and predicting answer quality [30]. However, the task of predicting the quality of physicians' answers in the HQA context has not been investigated. In this study, the similarities and differences between CQA and HQA were systematically compared, revealing that many non-textual features that are applicable to CQA answer quality prediction, especially those tightly coupled to the CQA webpage structure, are not applicable to HQA answer quality prediction. However, some general non-textual features unrelated to the webpage structure from previous studies of CQA answer quality prediction can be reused [17, 20, 26-30, 41-43], and based on the nature of HQA services, additional novel features are proposed. Based on the attributes of these collected features, they can be divided into two sets: surface linguistic features and social features. The surface linguistic features include features based on answer statistics, such as answer length, word density, average sentence length, and high-frequency domain words. The surface linguistic features also include features based on QA statistics, which describe the relationship between a question and its corresponding answer, such as the length ratio of the question and the answer, the word overlap in the QA pair and the similarity between the question and the answer. The social features include the statistical attributes of the profiles of the physicians participating in the HQA service. The general social features collected in this study include the time gap between question and answer, the overall number of patients, the overall number of visits, the personal grade of the physician, and the grade of the hospital at which the physician is employed. Additionally, several domain-specific social features were collected, such as the number of gifts received from patients, the number of telephone consultations and the overall number of registered patients after diagnosis. A complete summary of the feature sets considered is given in the appendix.

## 4. Results and discussion

### 4.1. Distribution of major non-textual features in the dataset

In this section, the ability of the non-textual features proposed in section 3.9 to predict high-quality answers in an HQA service is analyzed in detail. For this purpose, the relation between each non-textual feature and the ratio of high-quality answers versus low-quality answers is calculated. Due to space limitations, the feature significance analysis result in section 4.2 and feature universality are combined to extract three features from the surface linguistic feature set and social feature set for analysis. The three features extracted from the surface linguistic feature set are as follows: Number of repeated words in a QA pair, Similarity between question and answer and Answer length (number of characters in an answer). The three features extracted from the social feature set are as follows: Total number of patients, Overall number of visits and Time gap between question and answer.

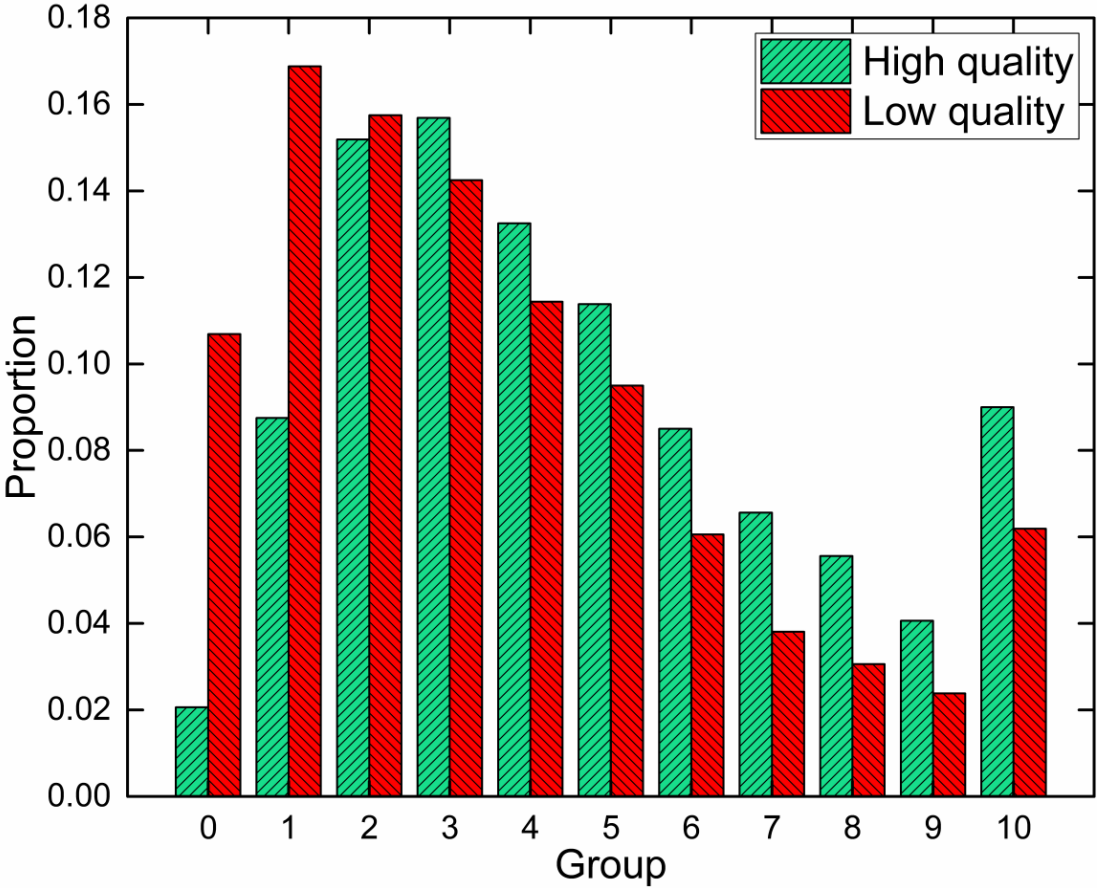

Fig. 3. Distribution of the "Number of repeated words in a QA pair" feature in the dataset.

As shown in Fig. 3, based on dataset statistics, the number of overlapping words in a QA pair is classified into 11 groups. Group 0 contains answers without any overlapping words in the question; group 1 contains answers with one overlapping word in the question; and so on. The 11$^{th}$ group contains answers with 10 or more overlapping words in the question. Fig. 3 shows that, as the number of overlapping words increases, the difference between the ratio of high-quality answers and the ratio of low-quality answers also increases. When the number of overlapping words is relatively small, the ratio of low-quality answers far exceeds the ratio of high-quality answers. By contrast, when the number of overlapping words is relatively large, the opposite is true. The main reason is that an answer with more overlapping words in the question is usually highly relevant to the question and contains complete information. Moreover, when the answer includes some words in the question, it not only makes the answer clearer but also shows that the answerer has given a careful answer to the question.

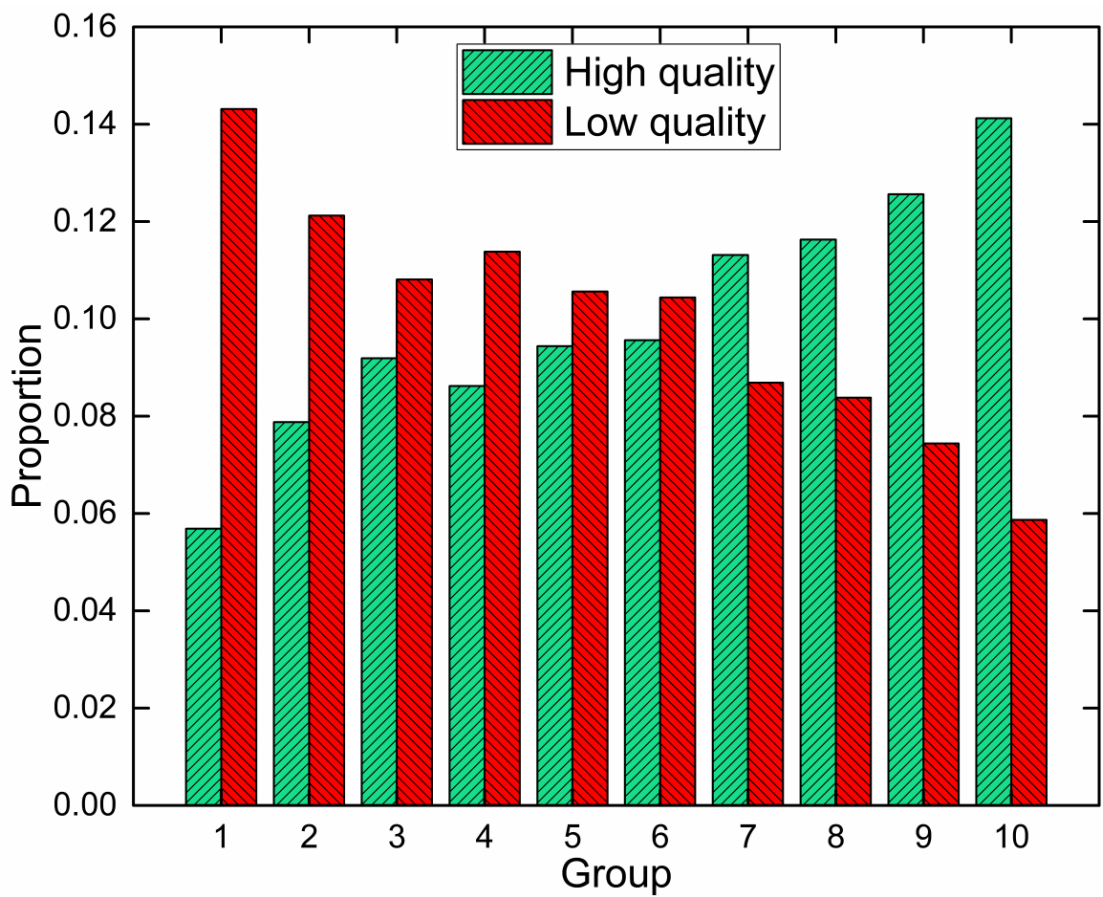

Fig. 4. Distribution of the "Similarity between question and answer" feature in the dataset.

Fig. 4 shows the recognition capability of the "Similarity between question and answer" feature. All answers are sorted in ascending order of similarity between QA pairs; then, they are divided into 10 groups with equal numbers, and the ratios of high-quality and low-quality answers in each group are calculated. As shown in the diagram, as similarity increases, the ratio of low-quality answers gradually decreases while the ratio of high-quality answers gradually increases. The main cause is that higher similarity between the QA pair means that the answer provided is more relevant to the question. An answer that is highly relevant to the question is normally a high-quality answer with abundant information.

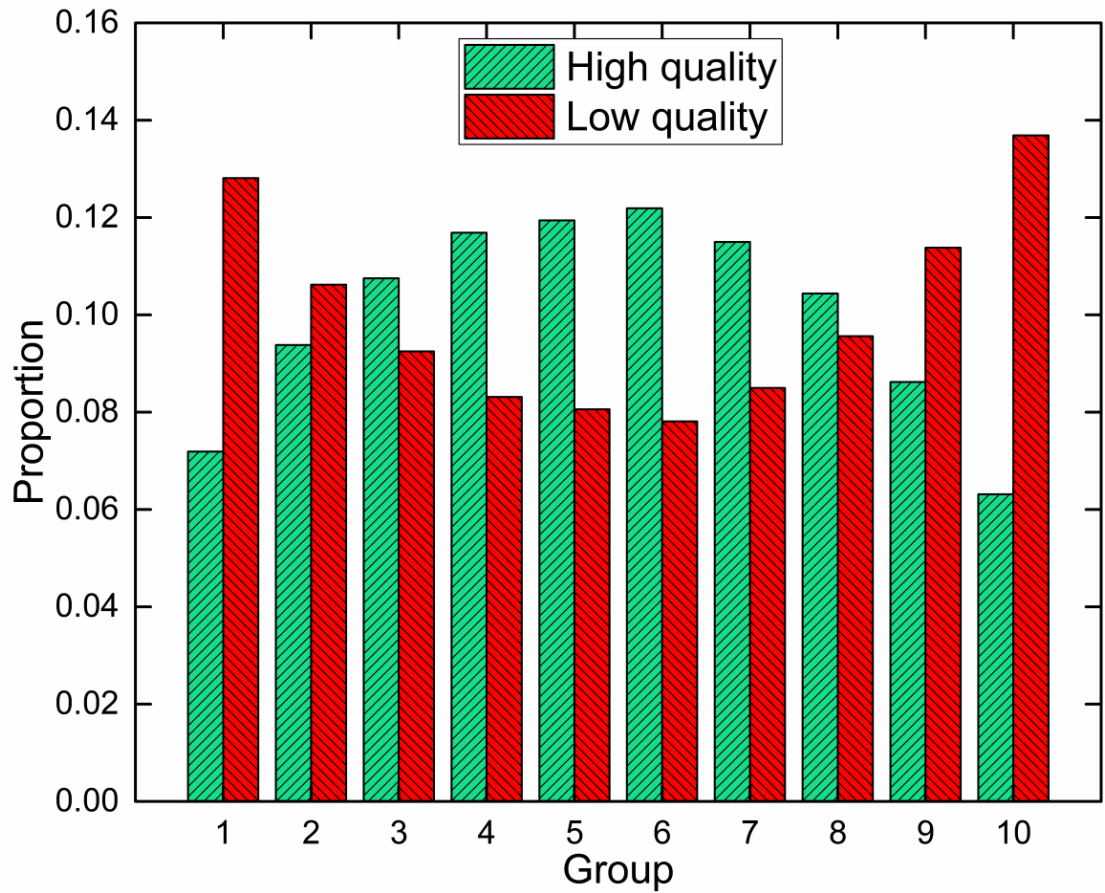

Fig. 5. Distribution of the "Answer length (number of characters in an answer)" feature in the dataset.

Fig. 5 describes the distribution of the "Answer length (number of characters in an answer)" feature in the dataset. All answers are sorted in ascending order of answer length and then divided into 10 groups to calculate the ratios of high-quality and low-quality answers in each group. As shown in the diagram, the ratio of high-quality answers initially increases with increasing answer length and then decreases, while the ratio of low-quality answers initially decreases with increasing answer length and then increases. To understand the reason for this behavior, the dataset was examined thoroughly. The results show that a very short answer is normally an irrelevant advertisement or greeting, while a very long answer is normally directly copied from another webpage such as Baidu Knows (http: //zhidao.baidu.com/) or Wikipedia (http: //www.wikipedia.org/). Therefore, we believe that the questioner prefers not to waste precious time on such answers and directly regards them as low-quality answers.

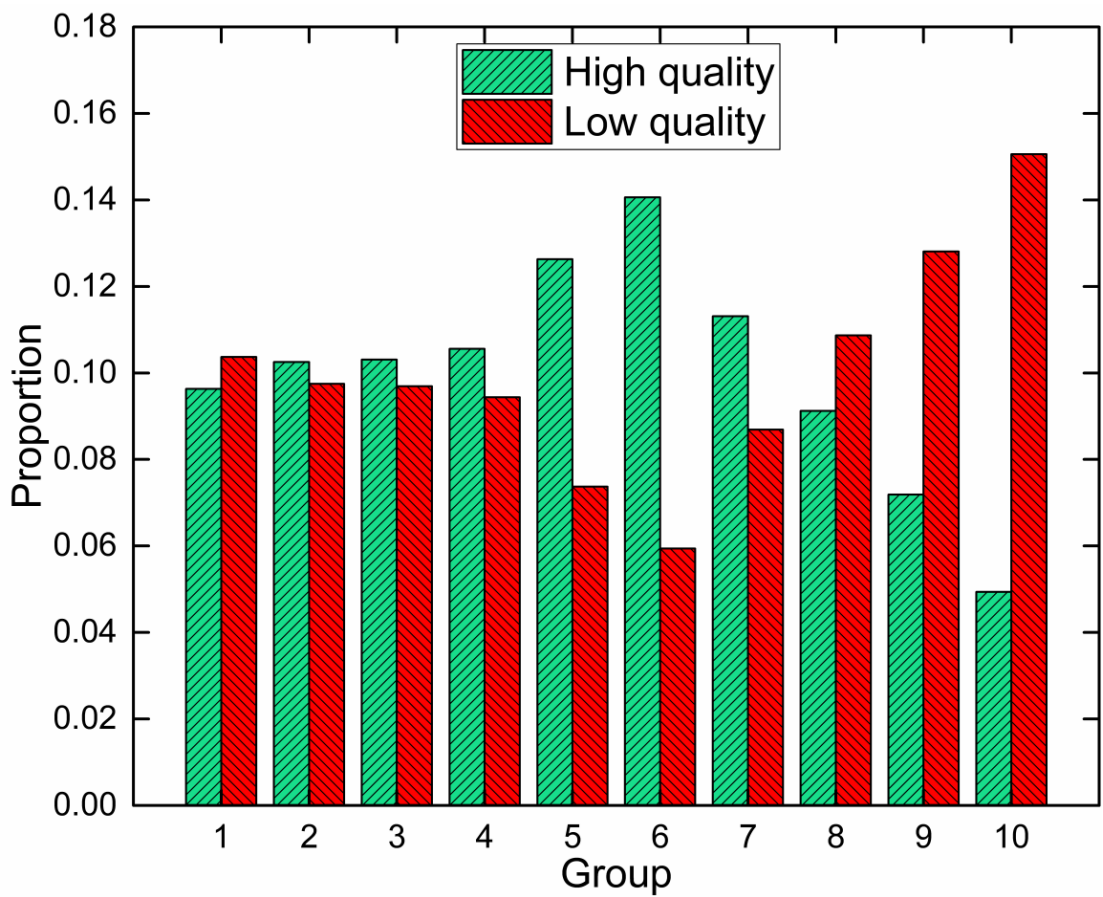

Fig. 6. Distribution of the "Total number of patients" feature in the dataset.

Fig. 6 shows the distribution of the "Total number of patients" feature in the dataset. All answers are sorted in ascending order according to the number of patients and then divided into 10 groups to calculate the ratios of high-quality and low-quality answers in each group. The diagram shows that the ratio of high-quality answers initially remains stable with increasing number of patients; then, it increases rapidly and finally decreases rapidly. The ratio of low-quality answer initially remains stable; then, it declines rapidly and finally increases rapidly. We believe the main reason is that the number of patients directly reflects a physician's popularity, and a more popular physician is normally more capable of providing a high quality-answer; however, it is worth noting that, as all physicians provide online consultation services for patient in their spare time, they are constrained by time and energy. Physicians with few patients might have adequate time, but the limitations of their own expertise mean that they may be unable to guarantee high-quality answers for each patient. Therefore, in the first group to the fourth group, the ratio of high-quality answers is close to the ratio of low-quality answers. When physicians have too many patients, even if they have required expertise, they have inadequate time and energy to provide high-quality answers to each patient. Therefore, in the $8^{th}$ group to the $10^{th}$ group, the ratio of high-quality answers decreases rapidly, while the ratio of low-quality answers increases rapidly. When the number of patients is moderate, physicians have both the required expertise and adequate time and energy to provide high-quality answers to each patient. Therefore, in the $5^{th}$ group to the $7^{th}$ group, the ratio of high-quality answers far exceeds the ratio of low-quality answers.

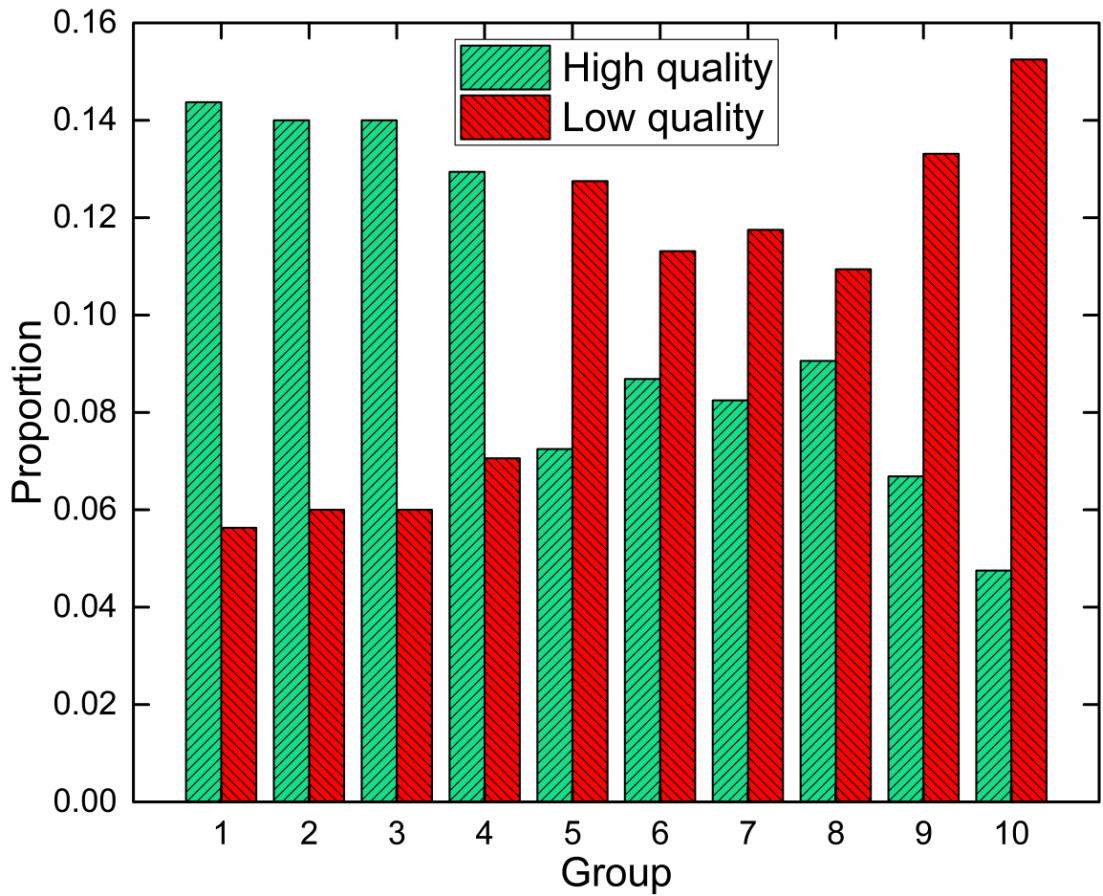

Fig. 7. Distribution of the "Overall number of visits" feature in the dataset.

Fig. 7 shows the distribution of the "Overall number of visits" feature in the dataset. All answers are sorted in ascending order of the total number of visits and then divided into 10 groups to calculate the ratios of high-quality and low-quality answers in each group. The diagram shows that in the $1^{st}$ group to the $4^{th}$ group, where the total number of visits is relatively low, the ratio of high-quality answers far exceeds the ratio of low-quality answers. By comparison, in the $9^{th}$ group to the $10^{th}$ group, where the total number of visits is relatively high, the ratio of low-quality answers far exceeds the ratio of high-quality answers. In the $5^{th}$ group to the $8^{th}$ group, where the total number of visits is moderate, even though the ratio of high-quality answers is close to the ratio of low-quality answers, overall, the former is smaller than the latter. We believe the main reason is that when the number of visits is relatively low, a physician has sufficient time and energy to provide high-quality answers to patient. By comparison, when the number of visits is extremely high, a physician has inadequate time and energy to provide high-quality answers to each patient.

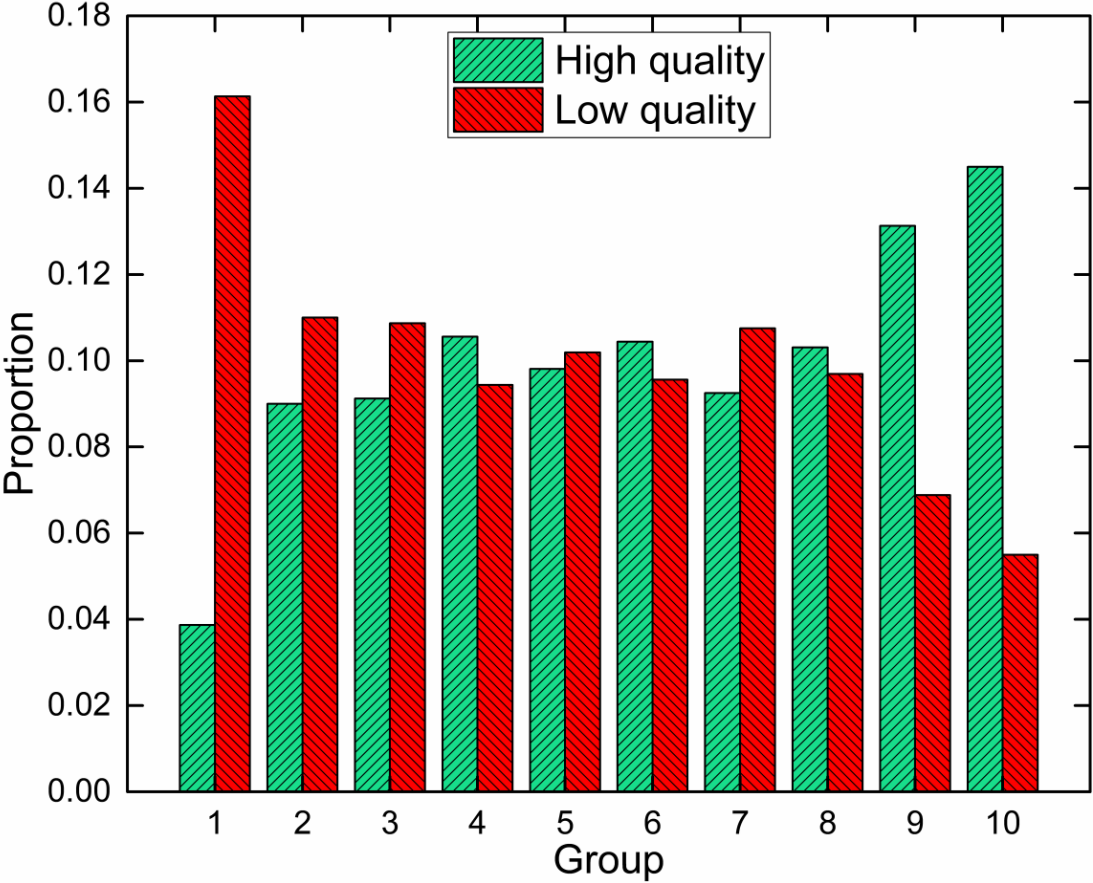

Fig. 8. Distribution of the "Time gap between question and answer" feature in the dataset.

Fig. 8 shows the distribution of the "Time gap between question and answer" feature in the dataset. All answers are sorted in ascending order of time difference and then divided into 10 groups to calculate the ratios of high-quality and low-quality answers in each group. The diagram shows that in the first group, which has the shortest response time, the ratio of low-quality answers far exceeds the ratio of high-quality answers. By contrast, in the 9th group to the 10th group, which have longer response times, the ratio of high quality answers far exceeds the ratio of low-quality answer. In the 2nd group to the 8th group, which have moderate response times, the ratio of high-quality answers is almost at the same level as the ratio of low-quality answers. To understand the reason, the dataset was examined carefully. The results show that in the group with the shortest response times, the answer provided by the physician is normally a template-based reply or advertisement. In the group with the longest response times, the answer provided by the physician is normally a complete and high-quality answer with a detailed explanation. In the group with moderate response times, the answers provided by the physician are of varying quality. The main reason is that a physician with a low level of expertise normally has adequate spare time to reply to a patient's question promptly. However, to improve their own popularity, they normally engage in various self-promotion activities. Physicians with high levels of expertise are normally busy and have only limited spare time to answer a patient's question online. Therefore, their response time is normally longer. However, they normally provide high-quality answers that are of great help to the patient.

The above analysis has proven that the proposed non-textual features can effectively distinguish high-quality answers from low-quality answers in an HQA service.

## 4.2. Analysis of feature importance

**Table 1**

Weight of each surface linguistic feature and social feature for the quality prediction task (feature names and definitions are given in section 3.9).

| Surface linguistic feature | Chi-squared statistic | Social feature | Chi-squared statistic |
|---|---|---|---|
| slf12 | 57.07433 | sf12 | 38.75335 |
| slf13 | 29.02399 | sf9 | 35.79734 |
| slf14 | 13.26076 | sf8 | 33.84864 |
| slf6 | 12.83025 | sf11 | 32.86179 |
| slf10 | 5.88354 | sf26 | 32.01754 |
| slf8 | 5.81761 | sf18 | 30.55206 |
| slf2 | 4.48754 | sf21 | 28.70678 |
| slf1 | 3.17281 | sf19 | 22.98488 |
| slf9 | 0.83498 | sf16 | 21.17314 |
| slf3 | 0.39188 | sf1 | 17.33798 |
| slf5 | 0.31550 | sf7 | 16.65631 |
| slf7 | 0.02832 | sf5 | 12.25441 |
| slf4 | 0.01495 | sf6 | 11.78288 |
| slf11 | 0.00566 | sf22 | 10.55804 |
| | | sf25 | 8.49262 |
| | | sf4 | 8.36756 |
| | | sf23 | 8.31638 |
| | | sf3 | 6.84316 |
| | | sf10 | 5.11407 |
| | | sf24 | 3.37639 |
| | | sf2 | 3.3719 |
| | | sf15 | 1.3807 |
| | | sf13 | 1.00437 |
| | | sf17 | 0.54971 |
| | | sf20 | 0.43875 |
| | | sf14 | 0.37992 |

When analyzing a classifier's performance, it is very useful to estimate the contribution of each feature. To calculate each feature's significance, the feature rank was evaluated based on chi-squared statistics [77], with the major features normalized to a range of 0 to 100. Table 1 lists the chi-squared statistics for the surface linguistic features and social features.

Table 1 shows that among all surface linguistic features defined in section 3.9, the number of repeated words in a QA pair (slf12) is the most prominent feature, followed by the number of repeated words in a QA pair (after the removal of stopwords) (slf13) and the similarity between a question and its corresponding answer (slf14). Our results differ from the findings of a previous work by Jeon et al., who reported that the answer length (slf1) was very important in the task of answer quality prediction [27]. By contrast, our results show that slf1 has no significant impact on answer quality prediction performance, whereas features that reflect the similarity between a question and its corresponding answer, such as slf12, slf13 and slf14, are more influential. In addition, the keyword density (slf6) is also a very effective feature for answer quality prediction. This result shows that in the HQA context, health consumers care greatly about the relevance and effectiveness of a physician's answer. Therefore, an answer that is closely related to the question and contains sufficient relevant information is typically regarded as a high-quality answer. By contrast, an answer with low relevance to the question, even if it is of sufficient length, is simply irrelevant and does not effectively solve the health consumer's problem; therefore, it is typically regarded as a low-quality answer by health consumers.

Among all of the social features listed in section 3.9, the total number of patients (sf12) is identified as the most effective feature, followed by the overall number of visits (sf9), the contribution value (sf8), and the total number of articles (sf11). The results show that a physician with higher values of sf12, sf9, sf8, and sf11 is more likely to provide high-quality answers. This is because a physician with a high sf12 value has accumulated extensive diagnosis experience, which attracts more health consumers for consultation and thus contributes to sf9; a higher value of sf8, which reflects the number of patients aided by a physician and the physician's activity on the website, indicates that the physician has helped more patients; and a higher sf11 value, which reflects a physician's professional level, corresponds to a physician with more extensive experience and knowledge who is therefore more likely to provide a trustworthy and professional high-quality answer.

*4.3. Analysis of classification performance*

The prediction of HQA answer quality is defined as a classical machine learning problem; therefore, the various textual features and non-textual features presented in this article were employed to train a LR model.

**Table 2**

Performances of the various sets of non-textual features (mean value ± standard deviation) with 5-fold cross validation (slf denotes the surface linguistic features, and sf denotes the social features).

| Feature Set | P (%) | R (%) | F1 (%) | AUC (%) |
|---|---|---|---|---|
| slf | **84.57±2.33** | **84.83±1.11** | **84.67±0.80** | **92.61±0.15** |
| sf | 73.69±2.78 | 73.90±2.94 | 73.76±2.19 | 83.84±2.28 |
| slf+sf | **88.98±2.88** | **89.12±1.34** | **89.03±1.59** | **95.81±0.53** |

First, the classification performances of the surface linguistic feature set and the social feature set introduced in section 3.9 were evaluated. Table 2 lists the results in terms of P, R, Fl, and AUC. As shown in Table 2, the set of surface linguistic features exhibits the best performance among all evaluation metrics. The primary reason for this superior performance is that the surface linguistic feature set consists of features based on answer statistics and QA pair statistics. These features provide an objective and fair evaluation of an answer's quality based on its content and its relevance to the corresponding question. By contrast, the social feature set consists of the statistical attributes of the profiles of the physicians participating in the HQA service, which reflect each physician's overall behavior and performance. However, a physician with generally excellent performance may provide low-quality answers under certain circumstances, for instance, if a health consumer's question is unclear or if the physician is too busy at that time to provide a comprehensive and detailed high-quality answer. By contrast, a physician with generally moderate performance may, for various reasons, occasionally provide a high-quality answer. For instance, such a physician may happen to specialize in the topic of a particular health consumer's question and also have sufficient spare time to provide a detailed and effective high-quality answer. Therefore, using the social feature set alone for answer quality prediction may lead to a certain degree of prediction deviation. In addition to the individual prediction performances of the non-textual feature sets, the combined performance of the surface linguistic feature set and the social feature set was investigated, as shown in the bottom row of Table 2. The findings show that the combination of the two non-textual features results in better accuracy than either non-textual feature set alone. These promising results demonstrate the effectiveness of the proposed surface linguistic features and social features.

**Table 3**

Performances of the various textual feature extraction methods (mean value ± standard deviation) with 5-fold cross validation.

| Feature Set | P (%) | R (%) | F1 (%) | AUC (%) |
| --- | --- | --- | --- | --- |
| Word (CHI-TFIDF) | 82.46±2.91 | 77.74±2.40 | 80.02±2.59 | 88.33±2.24 |
| Topic | 80.25±1.47 | 87.07±2.25 | 83.49±0.87 | 89.93±1.30 |
| Word (binary scheme) | 84.87±2.67 | 85.71±3.17 | 85.22±1.07 | 92.06±0.57 |
| DBN | **90.36±1.65** | **96.26±1.46** | **93.21±1.01** | **97.68±0.91** |

Table 3 lists the results obtained from the dataset via three classical textual feature extraction methods and the proposed deep learning approach. In Table 3, Word (binary scheme) corresponds to the first baseline method described in section 3.8, Word (CHI-TFIDF) corresponds to the second baseline method described in section 3.8, Topic denotes the third baseline method described in section 3.8, and DBN represents the proposed DBN-based deep learning approach, which is the fourth baseline method listed in section 3.8. These four textual feature extraction methods were all treated as baseline methods in the feature combination test presented in Table 5 and Table 6. Table 3 shows that the proposed DBN method, as expected, outperforms the previous three baseline methods in all evaluation metrics. Compared with the Word (CHI-TFIDF), Topic, and Word (binary scheme) methods, the performance of the proposed DBN method is, respectively, improved by 9.58%, 12.60%, and 6.47% in P, by 23.82%, 10.55%, and 12.31% in R, by 16.48%, 11.64%, and 9.38% in Fl, and by 10.59%, 8.62%, and 6.10% in AUC. These results provide sufficient proof of the effectiveness of the proposed method. The main reason for this performance improvement is that the DBN-based method learns a high-level semantic representation from the answers in the training set, thereby effectively overcoming the problem of feature sparsity in short text. The reason for the poor performance of the conventional textual feature extraction methods was also investigated and was found to be essentially attributable to the extracted HQA corpus. As previously mentioned, HQA content is typically in the form of short texts, resulting in feature sparsity.

Moreover, when the Word (CHI-TFIDF) method is used to extract textual features, both the keyword frequency and the contributions of keywords in each category are considered. When this rigorous feature extraction method is applied to short texts, many data samples can provide only limited features, and some samples may yield no features at all. This situation exacerbates the feature sparsity problem, and consequently, the Word (CHI-TFIDF) method exhibits the worst performance in all evaluation metrics. When the Topic method is used to extract textual features, the data samples are represented at a coarser granularity; thus, a more powerful representation of the data samples is obtained, which effectively alleviates the feature sparsity to some extent. Therefore, compared with the Word (CHI-TFIDF) method, the Topic method performs better in all evaluation metrics. Finally, when the Word (binary scheme) method is used to extract textual features, only the keyword frequency is considered. Although this simple feature extraction method does not consider the keyword contributions in each category, in the case of short texts, it somewhat alleviates the feature sparsity problem and consequently outperforms the Topic method in terms of the average number of features extracted from all data samples. Therefore, compared with the Word (CHI-TFIDF) and Topic methods, the Word (binary scheme) method performs better in all evaluation metrics.

**Table 4**

*p*-values of the T-test computed for various textual feature extraction methods over the Haodf dataset.

| Methods | Word (CHI-TFIDF) | Topic | Word (binary scheme) |
|---|---|---|---|
| DBN(P(%)) | 7.48027*1.0e-4 | 7.14105*1.0e-6 | 4.48115*1.0e-3 |
| DBN(R(%)) | 4.35866*1.0e-7 | 5.90379*1.0e-5 | 1.42241*1.0e-4 |
| DBN(F1(%)) | 5.52398*1.0e-6 | 2.07450*1.0e-7 | 1.96896*1.0e-6 |
| DBN(AUC(%)) | 2.47560*1.0e-5 | 4.33686*1.0e-6 | 2.58311*1.0e-6 |

To verify the reliability of the proposed deep learning approach, the T-test [78] was performed for the test results in Table 3. The *p*-values are listed in Table 4. All *p*-values are under 0.01, which shows that the proposed deep learning approach is significantly superior to the baseline method, again proving the reliability and effectiveness of the proposed method.

**Table 5**

Performances of the combined sets of features for the Word (CHI-TFIDF) and Topic methods (mean value ± standard deviation) with 5-fold cross validation (slf denotes the surface linguistic features, and sf denotes the social features).

| Feature Set | Word (CHI-TFIDF) | | | | Topic | | | |
|---|---|---|---|---|---|---|---|---|
| | P (%) | R (%) | F1 (%) | AUC (%) | P (%) | R (%) | F1 (%) | AUC (%) |
| Baseline | 82.46±2.91 | 77.74±2.40 | 80.02±2.59 | 88.33±2.24 | 80.25±1.47 | 87.07±2.25 | 83.49±0.87 | 89.93±1.30 |
| +slf | 89.03±1.68 | 92.24±1.00 | 90.60±1.29 | 96.46±0.59 | 89.55±1.02 | **93.76±1.12** | 91.60±0.48 | 97.03±0.68 |
| +sf | 84.22±2.33 | 88.82±1.84 | 86.45±1.85 | 93.31±1.19 | 83.68±2.43 | 89.57±1.34 | 86.51±1.42 | 93.43±1.09 |
| +slf+sf | **92.48±1.08** | **93.63±0.80** | **93.04±0.26** | **97.90±0.25** | **92.02±1.75** | 93.71±1.37 | **92.85±1.16** | **97.80±0.45** |

**Table 6**

Performances of the combined sets of features for the Word (binary scheme) and DBN methods (mean value ± standard deviation) with 5-fold cross validation (slf denotes the surface linguistic features, and sf denotes the social features).

| Feature Set | Word (binary scheme) | | | | DBN | | | |
|---|---|---|---|---|---|---|---|---|
| | P (%) | R (%) | F1 (%) | AUC (%) | P (%) | R (%) | F1 (%) | AUC (%) |
| Baseline | 84.87±2.67 | 85.71±3.17 | 85.22±1.07 | 92.06±0.57 | 90.36±1.65 | 96.26±1.46 | 93.21±1.01 | 97.68±0.91 |
| +slf | 90.49±1.93 | **92.74±1.72** | 91.58±0.92 | 96.27±0.34 | 94.92±0.82 | **97.83±0.94** | 96.35±0.59 | 98.98±0.52 |
| +sf | 86.44±3.05 | 87.51±2.54 | 86.92±1.38 | 93.37±0.65 | 92.42±1.46 | 96.27±1.63 | 94.29±0.99 | 97.97±0.72 |
| +slf+sf | **91.66±1.39** | 92.53±1.99 | **92.08±0.90** | **96.73±0.37** | **95.73±1.07** | 97.77±1.29 | **96.73±0.77** | **99.14±0.45** |

The combined performances of the textual feature sets and non-textual feature sets are evaluated. Table 5 and Table 6 list the results for various combinations of features. As shown in Table 5 and Table 6, the textual features extracted by the four textual feature extraction methods were used as baseline methods, and the various non-textual feature sets were added to these baselines. The table shows that when any set of non-textual features is added to any of the first three sets of textual features, the prediction performance somewhat improves. Furthermore, the addition of the surface linguistic feature set to any of the first three textual feature sets results in a more significant performance improvement in all evaluation metrics than does the addition of the social feature set. The main reason for this result is that the textual feature sets extracted by the three conventional textual feature extraction methods and the surface linguistic feature set complement each other better.

It is worth noting that when either of the two sets of non-textual features alone is added to the set of textual features extracted by the proposed DBN-based method, the overall performance shows no significant improvement. The primary reason for this finding may be that this fourth set of textual features is incompatible with either the surface linguistic feature set or the social feature set; thus, the addition of either of these feature sets may cause the performance to decline to some extent. However, the addition of both non-textual feature sets to each textual feature set results in a significant performance improvement; additionally, the results for all evaluation metrics except R are superior to the result obtained by adding only one non-textual feature set to a textual feature set. For instance, when both the surface linguistic feature set and social feature set are added to the textual feature set, in terms of the four evaluation metrics P, R, Fl, and AUC, the performance of the Word (CHI-TFIDF) method improves by 12.15%, 20.44%, 16.27%, and 10.83%, respectively; the performance of the Topic method improves by 14.67%, 7.63%, 11.21%, and 8.75%, respectively; the performance of the Word (binary scheme) method improves by 8.00%, 7.96%, 8.05%, and 5.07%, respectively; and the performance of the DBN method improves by 5.94%, 1.57%, 3.78%, and 1.49%, respectively. It is worth noting that the addition of both non-textual features to the proposed DBN method results in insignificant performance improvement. The main reason for this minimal improvement is that the presented DBN method is already able to achieve quite a high prediction performance based on the textual features alone such that any significant additional performance improvement would be difficult. This finding again demonstrates the effectiveness of the proposed method. Furthermore, when combined with the surface linguistic feature set and the social feature set, the proposed DBN method exhibits the best performance in all four prediction performance evaluation metrics except R, with P, R, Fl, and AUC values of 95.73%, 97.77%, 96.73%, and 99.14%, respectively. These test results show that the proposed DBN-based answer quality prediction framework and the proposed novel features are effective for predicting the quality of HQA answers.

## 5. Conclusions

To summarize, this research represents an initial effort to evaluate the quality of physicians' answers provided by HQA services. That is, several novel non-textual features, including surface linguistic features and social features, are combined with deep learning-based textual features with high-order semantic information to form a unified representation, which is used as input for a classification system. The DBN-based deep learning architecture has learnt an abstract high-order semantic representation from the HQA service dataset, which effectively overcomes the data sparsity problem that is often encountered in short text answer modeling and classification for HQA services. Several novel non-textual features reflect the HQA service answer quality from multiple angles such as answer content, the relation between a QA pair and physician profile statistics. The results from numerous tests prove that the proposed method and features can effectively predict HQA service answer quality.

Future research will focus on three further aspects of the problem. First, the proposed framework will be extended to address other problems in HQA, for instance, the prediction of the quality of health consumers' questions and the prediction of health consumers' levels of satisfaction. Furthermore, more novel effective features will be explored to further promote the prediction ability of the proposed method. Finally, a semi-supervised learning method will be introduced to address the problem of inadequately labeled data and to thus mitigate the heavy workload imposed by the need for manual labeling.

## Acknowledgements

This research was sponsored by the National Natural Science Foundation of China (No. 61370085), the National High Technology Research and Development Program of China (863 Program) (No. 2013AA01A205) and the Fundamental Research Funds for the Central Universities (No. HIT.NSRIF.2014067).

## Appendix: A complete summary of the non-textual features in section 3.9

- Surface linguistic features (14 in total)
    - Answer length (number of characters in an answer). This feature can be extracted easily without further answer analysis. A previous study has shown that this feature exhibits excellent performance in answer quality prediction [27]. (slf1)

- Number of words in an answer after tokenization. (slf2)

- Number of words in an answer after segmentation (removal of stopwords). (slf3)

- Number of non-repeated words in an answer. This feature represents the number of non-repeated words in an answer after segmentation and the removal of stopwords. (slf4)

- Character-word ratio in an answer after tokenization. This feature represents the ratio of the number of characters to the number of words in an answer after segmentation and the removal of stopwords. (slf5)

- Keyword density. This feature reflects the professionalism of a physician's answer. A physician's answer is typically a short text document, and it is expected that a good answer will contain as much useful information as possible within its limited length. (slf6 = keyword characters/total characters in answer)

- Number of sentences. This feature represents the number of sentences in an answer. A good answer is expected to contain an adequate number of sentences. (slf7)

- Average sentence length. This feature represents the average length of each sentence in an answer. The sentences in a good answer are expected to be of a reasonable length, i.e., they will contain sufficient information without being redundant. (slf8 = slf1/slf7)

- Number of high-frequency domain words in an answer. The number of high-frequency domain words reflects the authority and professionalism of a physician's answer. It is expected that a good answer will contain as many high-frequency domain words as possible within its limited length. In this study, 206293 terms were collected to compile a medical dictionary. Based on the online health question-answering service corpus, 1831 high-frequency domain words were extracted. (slf9)

- Ratio of question length to answer length. This feature provides the ratio of the length of a problem description to the length of the corresponding answer. It is expected that a good answer will be of sufficient length to contain enough detailed information to support its credibility. (slf10 = question length/slf1)

- Ratio of question length versus answer length (after removal of stopwords). (slf11 = question length(after stopword removal)/slf1(after stopword removal))

- Number of repeated words in a QA pair. This feature reflects the correlation between a patient's question and the physician's answer. A good answer is expected to be closely correlated with the corresponding question. (slf12)

- Number of repeated words in a QA pair (after the removal of stopwords). (slf13)

- Similarity between question and answer. This feature represents the similarity between a question and the corresponding answer. It is expected that a good answer will contain a reasonable ratio of the same terms used in the question. (slf14)

● Social features (26 in total)

- Time gap between question and answer. This feature reflects the promptness of a physician's reply. A good answer is expected to have a short response time. (sf1)

- Service rating after diagnosis. This feature reflects a physician's rating. In general, a higher rating indicates that a physician's answer is of higher quality. (sf2)

- Patient recommendation. This feature reflects a physician's popularity. A physician who provides high-quality answers is more likely to be recommended by patients. (sf3)

- Number of messages of thanks. This feature reflects the quality of a physician's answer. Only an answer of high quality is likely to receive messages of thanks from patients. (sf4)

- Number of gifts. This feature reflects a physician's answer quality. Only a physician who provides high-quality answers is likely to receive gifts from patients out of gratitude. (sf5)

- Number of gift givers. This feature reflects a physician's answer quality. A larger number of gift givers indicates that the answers provided by this physician are of higher quality. (sf6)

- Care value. This value increases in ratio to the number of gifts received by a physician and represents patients' gratitude to this physician. A higher care value indicates that the answers provided by this physician are of higher quality. (sf7)
- Contribution value. This feature represents the number of patients aided by a physician and that physician's activity on the website. A higher contribution value indicates that the answers provided by this physician are of higher quality. (sf8)
- Overall number of visits. This feature represents the overall number of patient visits to a physician's personal website from the moment when the physician first joined the online health question-answering service until the moment when the data were collected. (sf9)
- Number of previous-day visits. This feature represents the overall number of patient visits to a physician's personal website during the day prior to data collection. (sf10)
- Total number of articles. This feature represents a physician's contribution to the dissemination of medical health knowledge. (sf11)
- Total number of patients. This feature represents the number of patients aided by a physician. (sf12)
- Total number of registered outpatients after diagnosis. This feature represents the number of outpatients who registered offline at a physician's hospital after online consultation. (sf13)
- Number of registered outpatients after previous-day diagnosis. This feature represents the number of outpatients who registered offline at a physician's hospital after online consultation during the day before data collection. (sf14)
- Number of WeChat-registered outpatients after diagnosis. This feature represents the number of outpatients who registered at a physician's hospital via WeChat after online consultation. (sf15)
- Patient votes. This feature represents the patients' satisfaction level with a physician's answer, including evaluations of the treatment effectiveness and service attitude. (sf16)

- Activity (collection time – last online time). This feature represents the time gap between the time of data collection and the last time a physician was online. A smaller time gap indicates that the physician is more active. (sf17)
- Seniority (joining time – website launch time). This feature represents the time gap between the moment when a physician first joined the online health question-answering service and the time at which the online health question-answering service was launched. A smaller time gap indicates that the physician has greater seniority. (sf18)
- Grade. This feature includes five grades in ascending order as follows: no grade, hospital physician, physician, associate chief physician and chief physician. A higher grade indicates that a physician is more qualified and professional. (sf19)
- Hospital grade. This feature includes seven grades in ascending order as follows: no grade, grade one, grade one first class, grade two, grade two first class, grade three and grade three first class. A higher grade indicates that a hospital is more qualified, with more professional physicians and more advanced medical equipment. (sf20)
- Education. This feature includes five grades in ascending order as follows: no grade, assistant, lecturer, associate professor and professor. A higher grade indicates that a physician is more professional and more likely to provide high-quality answers. (sf21)
- Telephone service. This feature indicates whether a physician provides consultation services via telephone. Telephone services provide patients with more prompt and comprehensive consultation than online question-answering services. (sf22)
- Level of satisfaction with effectiveness of telephone consultation. This feature represents patients' satisfaction with the actual effectiveness of the telephone consultation services provided by a physician. Good answers are expected to be associated with a higher level of satisfaction. (sf23)

- Number of telephone consultations. This feature represents the number of telephone consultations requested from a physician by patients. A physician who provides high-quality answers is expected to be more likely to receive telephone consultation requests from patients. (sf24)
- Level of satisfaction with telephone consultation attitude. This feature represents patients' satisfaction with a physician's service attitude when providing telephone consultation services. Good answers are expected to be associated with a higher level of attitude satisfaction. (sf25)
- Number of direct communications with a physician via QR code scan. This feature represents the number of patients who have used smartphones or other mobile terminals for health consultations. A larger number of such communications indicates that a physician is more popular among patients. (sf26)

# References

[1] Y. Yin, Y. Zhang, X. Liu, Y. Zhang, C. Xing, H. Chen, HealthQA: A Chinese QA Summary System for Smart Health, in: X. Zheng, D. Zeng, H. Chen, Y. Zhang, D.B. Neill (Eds.) Smart Health, (Springer International Publishing, Cham, Switzerland, 2014), pp. 51-62.

[2] D. Dahlem, D. Maniloff, C. Ratti, Predictability bounds of electronic health records, Sci. Rep., 5 (2015) 11865.

[3] P. Klemm, K. Reppert, A nontraditional cancer support group: the Internet, Comput. Nurs., 16 (1998) 31-36.

[4] M.D. White, Questioning behavior on a consumer health electronic list, Libr. Q., 70 (2000) 302-334.

[5] J. Preece, Empathic communities: reaching out across the Web, Interactions, 5 (1998) 32-43.

[6] L.A. Slaughter, D. Soergel, T.C. Rindflesch, Semantic representation of consumer questions and physician answers, Int. J. Med. Inform., 75 (2006) 513-529.

[7] S.A. Adams, Blog-based applications and health information: two case studies that illustrate important questions for Consumer Health Informatics (CHI) research, Int. J. Med. Inform., 79 (2010) e89-e96.

[8] J. Frost, M. Massagli, Social uses of personal health information within PatientsLikeMe, an online patient community: what can happen when patients have access to one another's data, J. Med. Internet Res., 10 (2008) e15.

[9] T. Ginossar, Online participation: a content analysis of differences in utilization of two online cancer communities by men and women, patients and family members, Health Commun., 23 (2008) 1-12.

[10] A. Beloborodov, A. Kuznetsov, P. Braslavski, Characterizing health-related community question answering, Advances in Information Retrieval, (Springer, 2013), pp. 680-683.

[11] S. Oh, Y.J. Yi, A. Worrall, Quality of health answers in social Q&A, Proc. Am. Soc. Info. Sci. Technol., 49 (2012) 1-6.

[12] Y. Zhang, Contextualizing consumer health information searching: an analysis of questions in a

social Q&A community, Proceedings of the 1st ACM International Health Informatics Symposium, (ACM2010), pp. 210-219.

[13] G. Eysenbach, J. Powell, M. Englesakis, C. Rizo, A. Stern, Health related virtual communities and electronic support groups: systematic review of the effects of online peer to peer interactions, Bmj, 328 (2004) 1166.

[14] A. Beloborodov, P. Braslavski, M. Driker, Towards Automatic Evaluation of Health-Related CQA Data, Information Access Evaluation. Multilinguality, Multimodality, and Interaction, (Springer, 2014), pp. 7-18.

[15] T. Wang, Z. Huang, C. Gan, On mining latent topics from healthcare chat logs, Journal of biomedical informatics, 61 (2016) 247-259.

[16] C. Shah, S. Oh, J.S. Oh, Research agenda for social Q&A, Lib. Inf. Sci. Res., 31 (2009) 205-209.

[17] Q. Tian, P. Zhang, B. Li, Towards predicting the best answers in community-based question-answering services, Proceedings of the Seventh International AAAI Conference on Weblogs and Social Media (AAAI2013).

[18] K. Arai, A.N. Handayani, Predicting quality of answer in collaborative Q/A community, Soc. Cult., 2 (2013) 21-25.

[19] B. Li, T. Jin, M.R. Lyu, I. King, B. Mak, Analyzing and predicting question quality in community question answering services, Proceedings of the 21st international conference companion on World Wide Web, (ACM2012), pp. 775-782.

[20] Y. Liu, J. Bian, E. Agichtein, Predicting information seeker satisfaction in community question answering, Proceedings of the 31st annual international ACM SIGIR conference on Research and development in information retrieval, (ACM2008), pp. 483-490.

[21] B.M. Silva, J.J. Rodrigues, I. de la Torre Díez, M. López-Coronado, K. Saleem, Mobile-health: a review of current state in 2015, Journal of biomedical informatics, 56 (2015) 265-272.

[22] M. Lee, J.J. Cimino, H.R. Zhu, C. Sable, V. Shanker, J.W. Ely, H. Yu, Beyond Information Retrieval-Medical Question Answering, AMIA 2006 Symposium Proceedings, (AMIA2006), pp. 470-473.

[23] Y. Cao, F. Liu, P. Simpson, L. Antieau, A. Bennett, J.J. Cimino, J. Ely, H. Yu, AskHERMES: an online question answering system for complex clinical questions, J. Biomed. Inform., 44 (2011) 277-288.

[24] B.L. Cairns, R.D. Nielsen, J.J. Masanz, J.H. Martin, M.S. Palmer, W.H. Ward, G.K. Savova, The MiPACQ clinical question answering system, AMIA Annu. Symp. Proc., 2011 (2011) 171-180.

[25] F. Liu, L.D. Antieau, H. Yu, Toward automated consumer question answering: Automatically separating consumer questions from professional questions in the healthcare domain, Journal of biomedical informatics, 44 (2011) 1032-1038.

[26] E. Agichtein, C. Castillo, D. Donato, A. Gionis, G. Mishne, Finding high-quality content in social media, Proceedings of the 2008 International Conference on Web Search and Data Mining, (ACM2008), pp. 183-194.

[27] J. Jeon, W.B. Croft, J.H. Lee, S. Park, A framework to predict the quality of answers with non-textual features, Proceedings of the 29th annual international ACM SIGIR conference on Research and development in information retrieval, (ACM2006), pp. 228-235.

[28] L. Hoang, J.-T. Lee, Y.-I. Song, H.-C. Rim, A model for evaluating the quality of user-created documents, Information Retrieval Technology, (Springer, 2008), pp. 496-501.

[29] B.M. John, A.Y.-K. Chua, D.H.-L. Goh, What makes a high-quality user-generated answer?, IEEE Internet Comput., 15 (2011) 66-71.

[30] C. Shah, J. Pomerantz, Evaluating and predicting answer quality in community QA, Proceedings of the 33rd international ACM SIGIR conference on Research and development in information retrieval, (ACM2010), pp. 411-418.

[31] G.E. Hinton, R.R. Salakhutdinov, Reducing the dimensionality of data with neural networks,

Science, 313 (2006) 504-507.

[32] G.E. Hinton, S. Osindero, Y.-W. Teh, A fast learning algorithm for deep belief nets, Neural Comput., 18 (2006) 1527-1554.

[33] R. Salakhutdinov, G.E. Hinton, Deep boltzmann machines, International conference on artificial intelligence and statistics2009), pp. 448-455.

[34] T. Jo, J. Hou, J. Eickholt, J. Cheng, Improving protein fold recognition by deep learning networks, Sci. Rep., 5 (2015) 17573.

[35] S. Wang, J. Peng, J. Ma, J. Xu, Protein secondary structure prediction using deep convolutional neural fields, Sci. Rep., 6 (2016) 18962.

[36] S. Zhou, Q. Chen, X. Wang, Fuzzy deep belief networks for semi-supervised sentiment classification, Neurocomput., 131 (2014) 312-322.

[37] T. Kuremoto, S. Kimura, K. Kobayashi, M. Obayashi, Time series forecasting using a deep belief network with restricted Boltzmann machines, Neurocomputing, 137 (2014) 47-56.

[38] W.-L. Zheng, J.-Y. Zhu, Y. Peng, B.-L. Lu, EEG-based emotion classification using deep belief networks, 2014 IEEE International Conference on Multimedia and Expo (ICME), (IEEE2014), pp. 1-6.

[39] X.-h. ZHAO, J.-f. MA, Modify the method of feature's weight in text classfication, Comput. Knowl. Technol., 36 (2009) 209.

[40] D.M. Blei, A.Y. Ng, M.I. Jordan, Latent dirichlet allocation, J. Mach. Learn. Res., 3 (2003) 993-1022.

[41] J.-T. Lee, Y.-I. Song, H.-C. Rim, Predicting the quality of answers using surface linguistic features, in: C.Y. Ock, J.Y. Byun, Y.D. Bi (Eds.) Sixth international conference onadvanced language processing and web information technology, 2007. ALPIT, (IEEE2007), pp. 111-116.

[42] Z.-M. Zhou, M. Lan, Z.-Y. Niu, Y. Lu, Exploiting user profile information for answer ranking in cqa, Proceedings of the 21st international conference companion on World Wide Web, (ACM2012), pp. 767-

774.

[43] M. Huang, Y. Yang, X. Zhu, Quality-biased Ranking of Short Texts in Microblogging Services, IJCNLP2011), pp. 373-382.

[44] J.-T. Lee, Y.-I. Song, H.-C. Rim, Predicting the quality of answers using surface linguistic features, Advanced Language Processing and Web Information Technology, 2007. ALPIT 2007. Sixth International Conference on, (IEEE2007), pp. 111-116.

[45] J. Zhang, M.S. Ackerman, L. Adamic, Expertise networks in online communities: structure and algorithms, Proceedings of the 16th international conference on World Wide Web, (ACM2007), pp. 221-230.

[46] X.-H. Phan, L.-M. Nguyen, S. Horiguchi, Learning to classify short and sparse text & web with hidden topics from large-scale data collections, Proceedings of the 17th international conference on World Wide Web, (ACM2008), pp. 91-100.

[47] X. Yan, J. Guo, Y. Lan, X. Cheng, A biterm topic model for short texts, Proceedings of the 22nd international conference on World Wide Web, (ACM2013), pp. 1445-1456.

[48] M. Sahami, T.D. Heilman, A web-based kernel function for measuring the similarity of short text snippets, Proceedings of the 15th international conference on World Wide Web, (ACM, Edinburgh, Scotland, 2006), pp. 377-386.

[49] G. Zhou, Y. Liu, F. Liu, D. Zeng, J. Zhao, Improving question retrieval in community question answering using world knowledge, Proceedings of the Twenty-Third International Joint Conference on Artificial Intelligence2013), pp. 2239-2245.

[50] M. Chen, X. Jin, D. Shen, Short text classification improved by learning multi-granularity topics, in: T. Walsh (Ed.) Proceedings of the Twenty-Second International Joint Conference on Artificial Intelligence, (AAAI Press2011), pp. 1776-1781.

[51] Y. Kim, Convolutional neural networks for sentence classification, arXiv preprint arXiv:1408.5882,

(2014).

[52] J.M. Kleinberg, Authoritative sources in a hyperlinked environment, Journal of the ACM (JACM), 46 (1999) 604-632.

[53] L. Page, S. Brin, R. Motwani, T. Winograd, The PageRank citation ranking: bringing order to the web. Technical Report, Stanford InfoLab, 1999).

[54] J. Cho, S. Roy, R.E. Adams, Page quality: In search of an unbiased web ranking,  Proceedings of the 2005 ACM SIGMOD international conference on Management of data, (ACM2005), pp. 551-562.

[55] X. Zhu, S. Gauch, Incorporating quality metrics in centralized/distributed information retrieval on the World Wide Web,  Proceedings of the 23rd annual international ACM SIGIR conference on Research and development in information retrieval, (ACM2000), pp. 288-295.

[56] J. Bian, Y. Liu, D. Zhou, E. Agichtein, H. Zha, Learning to recognize reliable users and content in social media with coupled mutual reinforcement,  Proceedings of the 18th international conference on World wide web, (ACM2009), pp. 51-60.

[57] Y. Cai, S. CHAKRAVARTY, Answer Quality Prediction in Q/A Social Networks by Leveraging Temporal Features, International Journal of Next-Generation Computing, 4 (2013).

[58] F.M. Harper, D. Raban, S. Rafaeli, J.A. Konstan, Predictors of answer quality in online Q&A sites, Proceedings of the SIGCHI Conference on Human Factors in Computing Systems, (ACM2008), pp. 865-874.

[59] P. Jurczyk, E. Agichtein, Discovering authorities in question answer communities by using link analysis,  Proceedings of the sixteenth ACM conference on Conference on information and knowledge management, (ACM2007), pp. 919-922.

[60] B. Dom, I. Eiron, A. Cozzi, Y. Zhang, Graph-based ranking algorithms for e-mail expertise analysis, Proceedings of the 8th ACM SIGMOD workshop on Research issues in data mining and knowledge discovery, (ACM2003), pp. 42-48.

[61] C.S. Campbell, P.P. Maglio, A. Cozzi, B. Dom, Expertise identification using email communications, in: D. Kraft, O. Frieder, J. Hammer, S. Qureshi, L. Seligman (Eds.) Proceedings of CIKM '03 12th International Conference on Information and Knowledge Management, (ACM2003), pp. 528-531.

[62] L. Hong, B.D. Davison, A classification-based approach to question answering in discussion boards, Proceedings of the 32nd international ACM SIGIR conference on Research and development in information retrieval, (ACM2009), pp. 171-178.

[63] G.E. Hinton, Training products of experts by minimizing contrastive divergence, Neural Comput., 14 (2002) 1771-1800.

[64] U. Fiore, F. Palmieri, A. Castiglione, A. De Santis, Network anomaly detection with the restricted Boltzmann machine, Neurocomputing, 122 (2013) 13-23.

[65] S. Kim, Y. Choi, M. Lee, Deep learning with support vector data description, Neurocomputing, 165 (2015) 111-117.

[66] G.E. Hinton, T.J. Sejnowski, Learning and relearning in Boltzmann machines, Parallel Distributed Process. Explor. Microstruct. Cogn., 1 (1986) 282-317.

[67] P. Smolensky, Information processing in dynamical systems: Foundations of harmony theory, (Colorado University at Boulder, Dept. of Computer Science1986).

[68] R. Salakhutdinov, G. Hinton, Semantic hashing, Int. J. Approx. Reason., 50 (2009) 969-978.

[69] Y. Bengio, P. Lamblin, D. Popovici, H. Larochelle, Greedy layer-wise training of deep networks, Adv. Neural Inf. Process. Syst., 19 (2007) 153.

[70] R.-E. Fan, K.-W. Chang, C.-J. Hsieh, X.-R. Wang, C.-J. Lin, LIBLINEAR: a library for large linear classification, J. Mach. Learn. Res., 9 (2008) 1871-1874.

[71] C.-C. Chang, C.-J. Lin, LIBSVM: a library for support vector machines, ACM Trans. Intell. Syst. Technol., 2 (2011) 27.

[72] S. Rendle, Factorization machines with libFM, ACM Trans. Intell. Syst. Technol., 3 (2012) 57.

[73] F. Pedregosa, G. Varoquaux, A. Gramfort, V. Michel, B. Thirion, O. Grisel, M. Blondel, P. Prettenhofer, R. Weiss, V. Dubourg, Scikit-learn: machine learning in Python, J. Mach. Learn. Res., 12 (2011) 2825-2830.

[74] Z.E. Xu, M. Chen, K.Q. Weinberger, F. Sha, From sBoW to dCoT marginalized encoders for text representation, Proceedings of the 21st ACM international conference on information and knowledge management, (ACM2012), pp. 1879-1884.

[75] G. Salton, C. Buckley, Term-weighting approaches in automatic text retrieval, Inf. Process. Manag., 24 (1988) 513-523.

[76] G.E. Hinton, A practical guide to training restricted boltzmann machines, Neural Networks: Tricks of the Trade, (Springer, 2012), pp. 599-619.

[77] J.E. Cohen, The distribution of the chi-squared statistic under clustered sampling from contingency tables, J. Am. Stat. Assoc., 71 (1976) 665-670.

[78] D.H. Johnson, The insignificance of statistical significance testing, The journal of wildlife management, (1999) 763-772.